\documentclass[aps,pra,twocolumn,showpacs,superscriptaddress,longbibliography]{revtex4-1}
\usepackage{graphicx}  
\usepackage{dcolumn}  
\usepackage{bm}        
\usepackage{amssymb}   
\usepackage{amsmath}   
\usepackage{subfigure} 
\usepackage{hyperref} 
\hypersetup{
    colorlinks=true,
    linkcolor=blue,
    citecolor=blue,      
    urlcolor=blue,
    bookmarks=true,
}

\begin{document}

\title{Reply to ``Comment on ``High-pressure phases of group-II difluorides: Polymorphism and
superionicity'' ''} 

\author{Joseph R.~Nelson} \email{jn336@cam.ac.uk}
 \affiliation{Theory of Condensed Matter Group,
  Cavendish Laboratory, J.~J.~Thomson Avenue, Cambridge CB3 0HE,
  United Kingdom} 
\author{Richard J.~Needs} \affiliation{Theory of Condensed Matter Group,
  Cavendish Laboratory, J.~J.~Thomson Avenue, Cambridge CB3 0HE,
  United Kingdom}
\author{Chris J.~Pickard} \affiliation{Department of Materials Science
  and Metallurgy, University of Cambridge, 27 Charles Babbage Road,
  Cambridge CB3 0FS, United Kingdom} \affiliation{Advanced Institute
  for Materials Research, Tohoku University, 2-1-1 Katahira, Aoba,
  Sendai, 980-8577, Japan} \vskip 0.25cm

\begin{abstract}
  Cazorla \textit{et al}.~[preceding comment] criticize our recent
  results on the high-$PT$ phase diagram of CaF$_2$
  [\href{https://doi.org/10.1103/PhysRevB.95.054118}{Phys.~Rev.~B
    \textbf{95}, 054118 (2017)}]. According to our analysis, Cazorla
  \textit{et al}.~have not converged their calculations with respect
  to simulation cell size, undermining the comment's conclusions about
  both the high-$T$ behaviour of the $P\overline{6}2m$-CaF$_2$
  polymorph, and the use of the QHA in our work. As such, we take this
  opportunity to emphasise the importance of correctly converging
  molecular dynamics simulations to avoid finite-size errors. We
  compare our quasiharmonic phase diagram for CaF$_2$ with currently
  available experimental data, and find it to be entirely consistent
  and in qualitative agreement with such data. Our prediction of a
  superionic phase transition in $P\overline{6}2m$-CaF$_2$ (made on
  the basis of the QHA) is shown to be accurate, and we argue that
  simple descriptors, such as phonon frequencies, can offer valuable
  insight and predictive power concerning superionic behaviour.
\end{abstract}

\vskip 0.25cm


\maketitle

Part of our recent work \cite{Nelson_PRB_2017} reports a
pressure-temperature phase diagram for CaF$_2$ (Fig.~8 of
\cite{Nelson_PRB_2017}) calculated using the quasiharmonic
approximation (QHA). This phase diagram was constructed by first
searching the potential energy surface of CaF$_2$ using the \textit{ab
  initio} random structure searching (AIRSS) technique \cite{AIRSS},
after which the effects of finite temperature were treated by
explicitly calculating the Gibbs free energy of candidate low-enthalpy
phases in the QHA. We found that in the QHA, the Gibbs free energy
difference between the known high-pressure $\gamma$-CaF$_2$ phase, and
a $P\overline{6}2m$-symmetry structure found using AIRSS, closes at
increased temperature, leading us to propose
\mbox{$P\overline{6}2m$-CaF$_2$} as a high-$T$ CaF$_2$ phase. We
consider this to be quite reasonable, given that the same
$P\overline{6}2m$ structure is already a known high-$T$ phase in other
alkaline earth metal halides \textit{e.g.}~BaCl$_2$ \cite{BaCl2_P62m}
and BaI$_2$ \cite{BaI2_P62m}.

We also proposed that $P\overline{6}2m$-CaF$_2$ would undergo a
superionic phase transition at still higher temperatures. This
conclusion was reached through the identification and analysis of an
unstable $K$-point phonon mode present in the $P\overline{6}2m$
structure (Figs.~10 and 11 of \cite{Nelson_PRB_2017}).

In reply to Cazorla and Errandonea's (hereafter `CE') criticism
\cite{Cazorla_Comment_2018}, we demonstrate in Sec.~\ref{PhaseDiagram}
that our CaF$_2$ phase diagram is completely consistent with available
experimental data on high-$PT$ CaF$_2$. In Sec.~\ref{AIMD}, we provide
the results of \textit{ab initio} molecular dynamics (AIMD)
simulations on $P\overline{6}2m$-CaF$_2$. These simulations fully
substantiate our prediction of a superionic phase transition in this
polymorph \cite{Nelson_PRB_2017}, and show that
$P\overline{6}2m$-CaF$_2$ is stable in one-phase simulations to at
least \mbox{3000 K} at \mbox{20 GPa}. We find that the AIMD
simulations presented by CE in \cite{Cazorla_Comment_2018} have not
used sufficiently large simulation cells; accordingly, CE's results on
high-$T$ $P\overline{6}2m$-CaF$_2$ are erroneous. Given that CE base
much of their criticism of our work on those flawed results, we are
led to conclude that their comment is without merit. We end our reply
with a discussion concerning phonons, the QHA, and superionicity.

\section{\label{PhaseDiagram}PT phase diagram of CaF$_2$}
In Fig.~\ref{fig:PhaseDiagram} below, we plot the boundaries of our
QHA phase diagram (blue lines $-$ Fig.~8 of \cite{Nelson_PRB_2017})
alongside experimental data from Refs.~\cite{Cazorla_Comment_2018}
(points with error bars) and \cite{Mirwald_JPCS_1980} (solid green
lines). An examination of Fig.~\ref{fig:PhaseDiagram} demonstrates
that our QHA results are entirely consistent with both sets of
experimental data. We identify two low-temperature solid states $-$
the known $\alpha$ and $\gamma$ phases $-$ and our $\alpha$-$\gamma$
phase boundary shows good quantitative agreement with the data of
\cite{Mirwald_JPCS_1980}, while the data of
Ref.~\cite{Cazorla_Comment_2018} lies about 3 GPa higher. Our QHA
results suggest a high-temperature modification of $\alpha$-CaF$_2$ as
thermal expansion drives this compound to volumes at which it exhibits
a harmonic phonon instability; this is consistent with
Refs.~\cite{Cazorla_Comment_2018,Mirwald_JPCS_1980} and the known
high-$T$ $\beta$-CaF$_2$ phase.

Experimentally, the behaviour of CaF$_2$ at higher pressures and
temperatures is less clear. For example, following the
$\alpha$-$\gamma$ boundary with increasing temperature, the data of
Ref.~\cite{Mirwald_JPCS_1980} suggests (based on changes in the slope
of the $\alpha$-$\gamma$ boundary) a possible triple point involving
the solid $\alpha$ and $\gamma$ phases, but not the superionic
$\beta$ phase, to occur first, and this is what we qualitatively
find in our QHA phase diagram. CE on the other hand have proposed that
a triple point involving the $\alpha$, $\beta$ and $\gamma$ phases
occurs first (Fig.~1(a) of
\cite{Cazorla_Comment_2018}). Quantitatively, these triple points are
separated by about \mbox{800 K}, with our QHA results half-way in
between (Fig.~\ref{fig:PhaseDiagram}). We remark here that the
experimental results in
Refs.~\cite{Cazorla_Comment_2018,Mirwald_JPCS_1980} do not offer any
structural information about high-$PT$ CaF$_2$ phases.

\begin{figure}
\centering
  \includegraphics[scale=0.43]{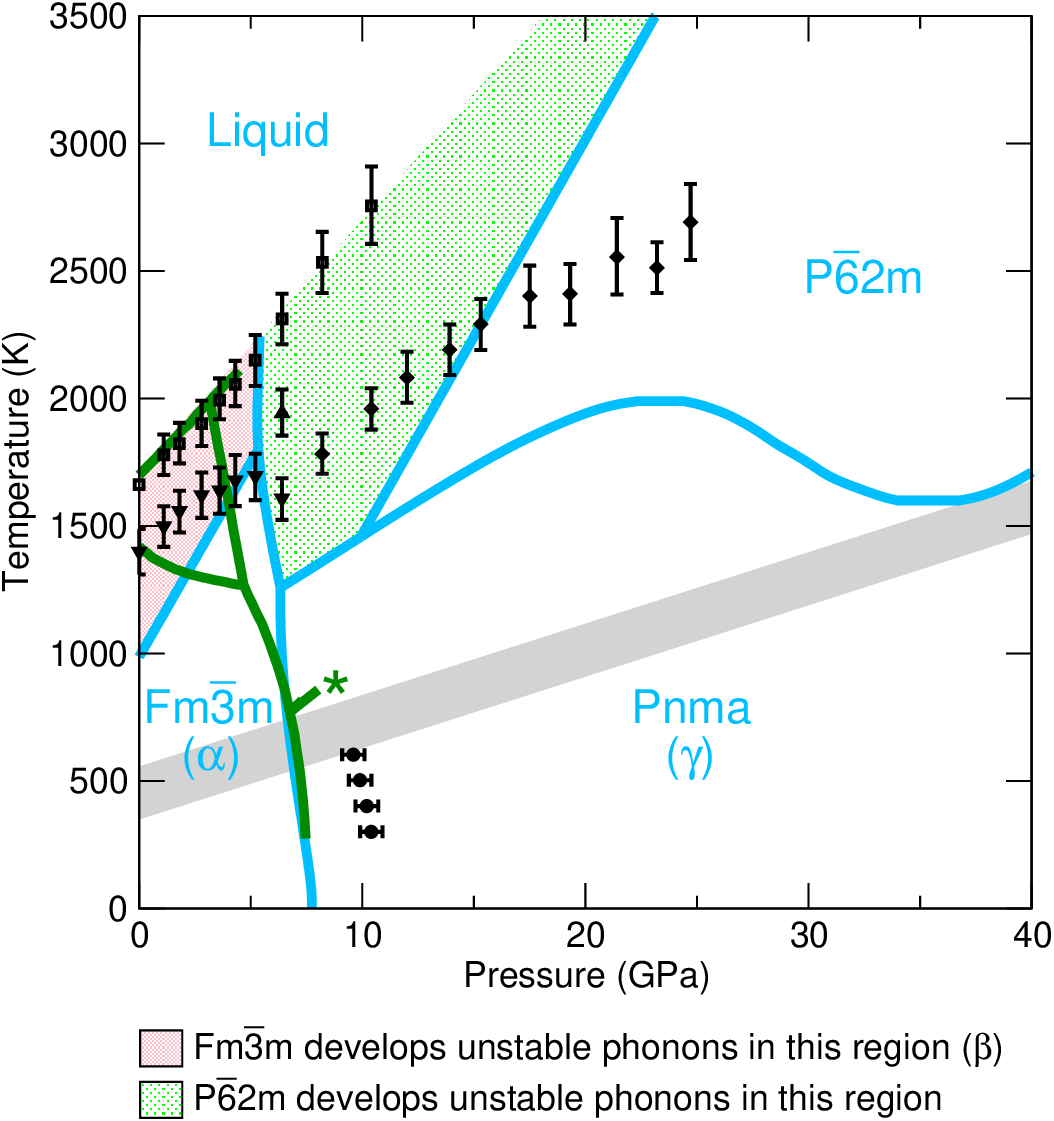}
  \caption{\label{fig:PhaseDiagram} Pressure-temperature phase
    diagrams for CaF$_2$. Blue lines are the QHA results of our work
    \cite{Nelson_PRB_2017}, points with errorbars are from
    \cite{Cazorla_Comment_2018}, and green lines are from
    \cite{Mirwald_JPCS_1980}. The green asterisk marks a possible
    triple point according to Ref.~\cite{Mirwald_JPCS_1980}. The
    light-grey band running diagonally up and right is where
    $d^2\alpha/dT^2$ changes sign in the QHA, and pertains to its
    validity, discussed in Sec.~\ref{Predict}.}
\end{figure}

Our QHA phase diagram uses a straight line to represent the melting
curve of CaF$_2$. This line is consistent with experimental data given
in Ref.~\cite{Cazorla_Comment_2018} (see
Fig.~\ref{fig:PhaseDiagram}). In practice, there is likely to be some
curvature or change in curvature in the melting curve, particularly at
any triple points.

% CE have, in their own previous work, also used a straight line to
% describe the melting curve of CaF$_2$ (Fig.~1 of
% \cite{Cazorla_arxiv}).

The coexistence of $\beta$-CaF$_2$ with another superionic phase
cannot be ruled out by currently available experimental data. As CE
correctly acknowledge in their comment, the phase with which
\mbox{$\beta$-CaF$_2$} coexists (excluding the $\alpha$ and liquid
phases) is ``experimentally not resolved''. Previous theoretical work
on the CaF$_2$ phase diagram is divided as to whether such a phase is
solid, superionic or even liquid.

CE describe our $\gamma$-$P\overline{6}2m$ phase boundary as
``inconsistent'' with the experimental data in
Ref.~\cite{Cazorla_Comment_2018}, however Fig.~\ref{fig:PhaseDiagram}
makes it obvious that this is not the case. Both sets of data (the
blue $\gamma$-$P\overline{6}2m$ phase boundary, and the points with
solid diamonds and errorbars in Fig.~\ref{fig:PhaseDiagram}) are in
clear qualitative agreement. The experimental data in this case does
not extend to sufficiently high pressures to ascertain whether this
boundary then falls in temperature, as predicted in the QHA. We
reiterate here that the calculation of this particular phase boundary
is subject to some uncertainty depending on the choice of
equation-of-state (EOS) used for the Gibbs free energy calculation;
this point, and details of the EOS we used for these calculations were
comprehensively addressed in our work \cite{Nelson_PRB_2017} so we
refer the interested reader there. We also remark that in the QHA, the
free energy surfaces of the $\gamma$ and $P\overline{6}2m$ phases are
almost parallel (Fig.~6 of \cite{Nelson_PRB_2017}), which introduces
uncertainty in calculating their high-$T$ intersections
\cite{Schieber_JCP_2018}.

In our QHA CaF$_2$ phase diagram, we would expect the boundaries
between the $\alpha$ and $P\overline{6}2m$ phases, and the regions in
which these phases develop phonon instabilities, to be essentially
linear (Fig.~\ref{fig:PhaseDiagram}). This is because these boundaries
are isochores, corresponding to \mbox{$V=46.47$
  \AA$^3$/f.u.}~($\alpha$-CaF$_2$) and \mbox{$V=37.70 $
  \AA$^3$/f.u.}~($P\overline{6}2m$-CaF$_2$) \cite{Nelson_PRB_2017},
and is unrelated to the QHA.

A comparison of CE's previous AIMD simulations on CaF$_2$ and our QHA
phase diagram is unproductive. CE's previous simulations are limited
to the $\alpha$ and $\gamma$ CaF$_2$ phases only, and do not consider
the $P\overline{6}2m$ CaF$_2$ structure proposed in our work
\cite{Nelson_PRB_2017}. Should a $T$-induced transition between the
$\gamma$ and $P\overline{6}2m$ phases occur, as we predicted using the
QHA, it is extremely unlikely that CE would have observed this
transition in their simulations: among the difficulties are that the
$P\overline{6}2m$ structure, with 9 atoms/unit cell, is incompatible
with the 192-atom supercells used in CE's previous work. The behaviour
of CaF$_2$ in those simulations should not therefore be expected to
mirror the results found in our QHA phase diagram. The only AIMD
simulations that CE do carry out on the $P\overline{6}2m$-CaF$_2$
structure have their own deficiencies, which we discuss in the next
Section.

\section{\label{AIMD}High temperature behaviour of
  $P\overline{6}2m-$CaF$_2$}

We suggested in Ref.~\cite{Nelson_PRB_2017} that
$P\overline{6}2m$-CaF$_2$ would become superionic at sufficiently high
temperatures. CE, however, present AIMD simulations in their comment
(Figs.~2(a) and (b) of \cite{Cazorla_Comment_2018}) in which they
claim that $P\overline{6}2m$-CaF$_2$ is not superionic.

We discuss Fig.~2(a) of \cite{Cazorla_Comment_2018} first, which shows
the results of an AIMD simulation on $P\overline{6}2m$-CaF$_2$ at
\mbox{$T=2500$ K} and \mbox{$P=20$ GPa}. Under these conditions,
according to CE, the Ca sublattice in $P\overline{6}2m$-CaF$_2$ melts
and Ca ions show diffusive behaviour. These results are unphysical:
the diffusion of Ca (as opposed to F) is not expected. Coulombic
arguments would instead lead us to expect that diffusion is more
energetically costly for Ca$^{2+}$ due to its higher ionic charge
relative to F$^{-}$ \cite{Hull_RPP_2004}, which is supported by direct
measurements of anion and cation diffusion coefficients in
$\alpha$-CaF$_2$ \cite{Matzke_JMS_1970}. As such, CE's apparent
observation of Ca diffusion suggests a problem with their AIMD
simulations, rather than a criticism of our work. We therefore present
our own results here. Our AIMD calculations use the \textsc{cp2k}
code, the PBE exchange-correlation functional, GTH pseudopotentials
for Ca and F, and DZVP `MOLOPT' Gaussian basis sets
\cite{CP2K,PBE1996,GTH_pseudos_1,GTH_pseudos_2,GTH_pseudos_3,MOLOPT}.
Mean-squared displacement (MSD) calculations average over both the
atoms of a particular species (Ca or F), and over different time
origins. Positions are corrected for the center-of-mass motion in all
cases.

\begin{figure}
\centering
  \includegraphics[scale=0.43]{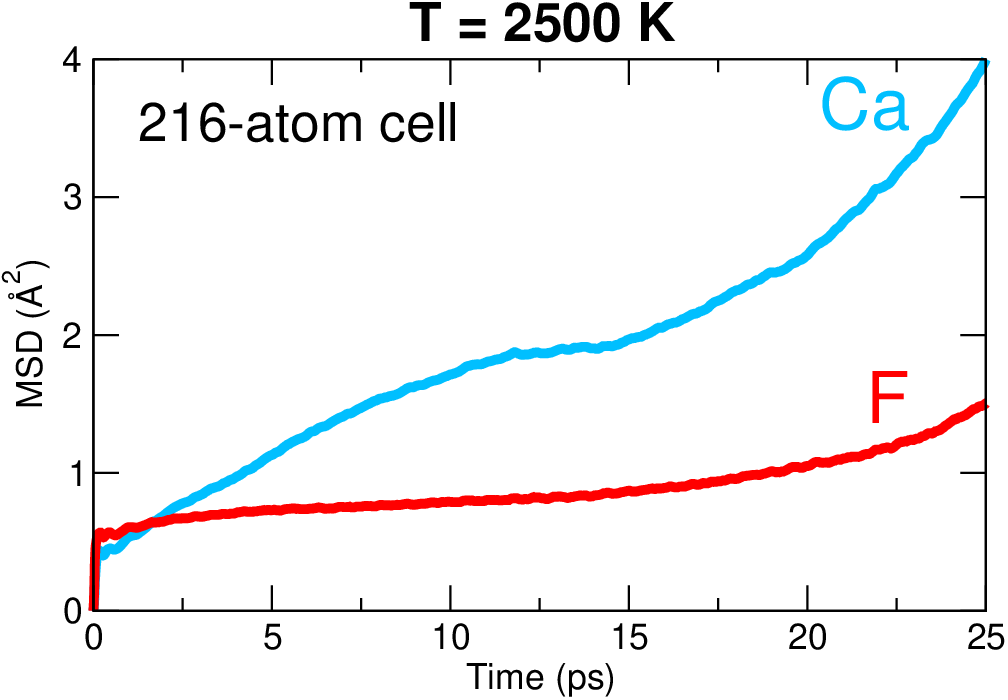}
  \caption{\label{fig:smallercell} Mean-squared displacement of F and
    Ca ions in $P\overline{6}2m$-CaF$_2$ in our 216-atom AIMD-$NVT$
    simulation at \mbox{$T=2500$ K}. The pressure is $20.0\pm 1.0$
    GPa.}
\end{figure}

Fig.~\ref{fig:smallercell} shows the MSD of Ca and F ions at
\mbox{$P=20$ GPa} and \mbox{$T=2500$ K} for $P\overline{6}2m$-CaF$_2$,
using a supercell containing 216 atoms, which is of a similar size to
that used by CE in their comment. Comparing
Figs.~\ref{fig:smallercell} and Fig.~2(a) of
\cite{Cazorla_Comment_2018}, we indeed find analogous results: there
is a relative stasis in F diffusion, and prominent Ca
diffusion. However, it is clear that this behaviour does not persist
upon increasing the simulation cell size. In
Ref.~\cite{Nelson_PRB_2018}, we show that large simulation cells are
required to obtain results that are qualitatively converged with
respect to simulation cell size. 

\begin{figure}
\centering
  \includegraphics[scale=0.28]{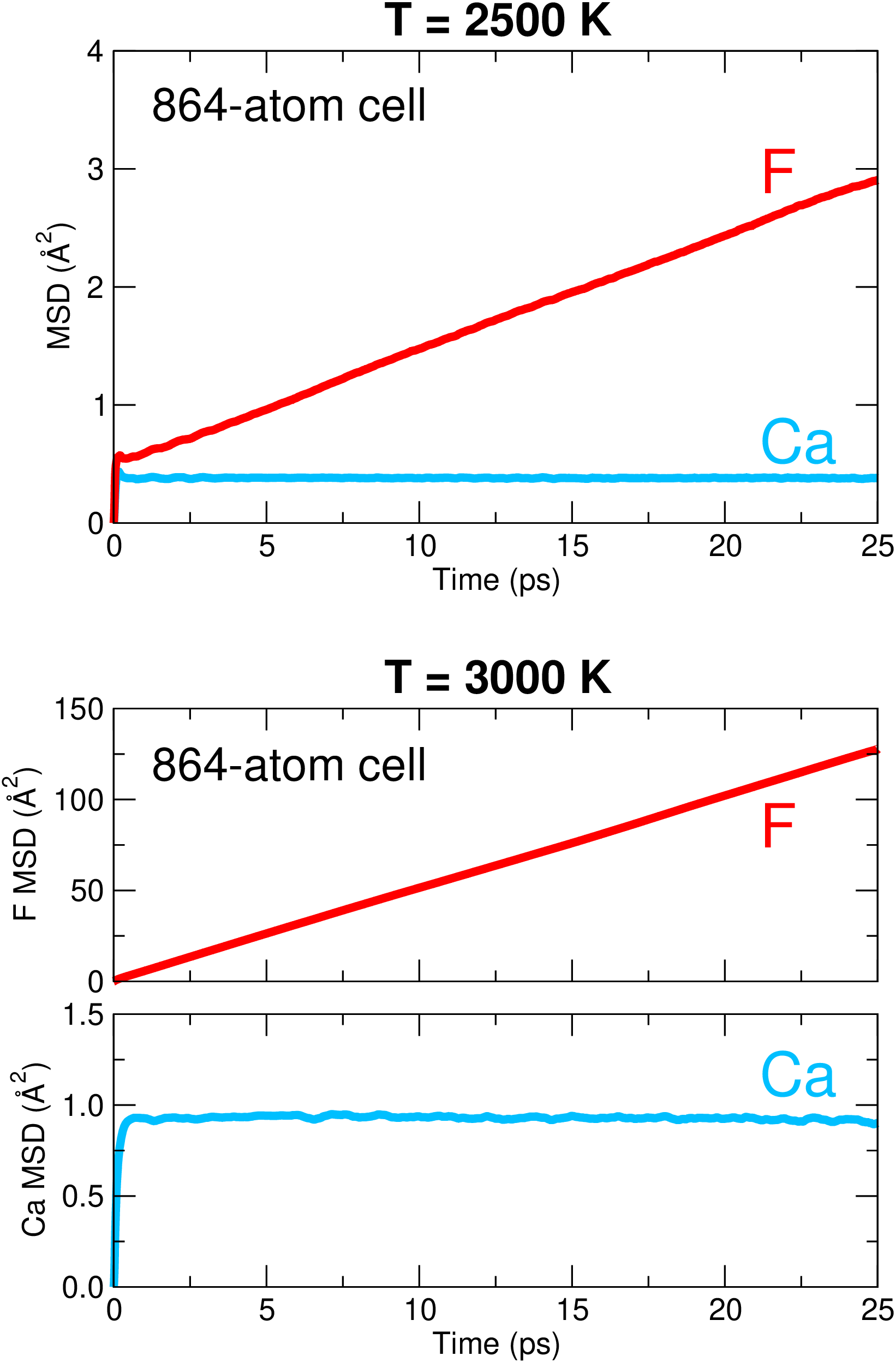}
  \caption{\label{fig:T_2500K_3000K} Mean-squared displacement of F
    and Ca ions in $P\overline{6}2m$-CaF$_2$ in our 864-atom
    AIMD-$NVT$ simulations at \mbox{$T=2500$ K} and \mbox{3000 K}. The
    pressure is \mbox{$P=19.9\pm0.4$ GPa} at both temperatures. These
    are to be compared directly to Fig.~2 of the comment
    \cite{Cazorla_Comment_2018}. Note that, in clear contrast to the
    claims made in \cite{Cazorla_Comment_2018}, there is no unphysical
    diffusion of Ca ions at either temperature, and that the system is
    not in a melt state at 3000 K. The high diffusion of F ions at
    3000 K is indicative of a superionic state.}
\end{figure}

Fig.~\ref{fig:T_2500K_3000K} provides the results of an AIMD
simulation carried out with a larger, 864-atom cell. The results we
obtain are very different from those using supercells with 216
atoms. At \mbox{$P=20$ GPa} and \mbox{$T=2500$ K}, we observe mild F
diffusion; note that there is no unphysical Ca diffusion. At
\mbox{$T=3000$ K} (with the same cell size), F diffusion is
significantly higher than at \mbox{$T=2500$ K}, consistent with a
superionic phase transition occurring in $P\overline{6}2m$-CaF$_2$
between these two temperatures. We do not find
$P\overline{6}2m$-CaF$_2$ to be in a melt state at \mbox{3000 K}.

These results support the conclusions of our work
\cite{Nelson_PRB_2017}: $P\overline{6}2m$-CaF$_2$ undergoes a
high-temperature superionic phase transition. A comparison of
Figs.~\ref{fig:smallercell} and \ref{fig:T_2500K_3000K} highlights the
importance of correctly converging MD simulations with respect to
supercell size in order to avoid drawing erroneous
conclusions. Finite-size effects are, in this case, so severe that
they led CE to predict the wrong diffusing species (Ca instead of F)
at \mbox{2500 K} and \mbox{20 GPa} \cite{Cazorla_Comment_2018}. On the
basis of the results in this section, we firmly reject CE's claims
that $P\overline{6}2m$-CaF$_2$ is not superionic, and that its melting
temperature is below 3000 K. We are led to conclude that
$P\overline{6}2m$-CaF$_2$ is an excellent candidate for a high-$PT$
CaF$_2$ phase.

\section{\label{Predict}Phonons and superionicity}
There is a considerable amount of literature discussing phonon modes
$-$ either soft, or low-energy $-$ in conjunction with superionic
conductivity
\textit{e.g.}~\cite{Boyer_PRL_1980,Boyer_Phase_1981,Samara_SSP_1984,Zhou_SSC_1996,Wakamura_PRB_1997,Wakamura_PRB_1999,Gupta_PRB_2012,Buckeridge_Ceria_2013,Ghosh_PCCP_2016,Roychowdhury_AChem_2018},
and phonon modes have also been invoked to explain other types of
diffusive behaviour, such as self-diffusion in transition metals
\cite{Herzig_BB_1989,Marian_PRB_2002}.

The softening of a phonon mode at the harmonic level results in an
increase in amplitude of the softening mode and the corresponding
creation of a double-well energy potential \cite{Boyer_Ferro_1990},
which can favour defect creation and increase the probability of
interstitial site occupation, as discussed for superionic ThO$_2$
\cite{Ghosh_PCCP_2016}. However, the details require analysis of the
soft mode's eigenvectors
\cite{Boyer_PRL_1980,Buckeridge_Ceria_2013,Ghosh_PCCP_2016,Nelson_PRB_2017}
to understand their effect on ionic motion. We consider this to be an
important point. Reading CE's comment, one is left with the
unfortunate impression that any soft phonon mode would result in a
superionic phase transition; this does not follow, since the soft mode
could be indicative of a different kind of transition,
\textit{e.g.}~structural, ferroelectric, and so on. Instead, the mode
(or combination of modes \cite{Buckeridge_Ceria_2013}) should be
examined to see whether they exhibit behaviour conducive to
superionicity, such as leaving one atomic species motionless
\cite{Boyer_PRL_1980}, or promoting the movement of ions toward
interstitial sites \cite{Buckeridge_Ceria_2013}. This is how we
proceeded in Ref.~\cite{Nelson_PRB_2017}.

Previous studies have used the QHA and/or phonon modes to explain
superionic behaviour in SrF$_2$, BaF$_2$, Li$_3$OCl and AgI
\cite{Boyer_SSI_1981,Makur_PSSB_1992,Zhao_JACS_2012,Buhrer_SSC_1975}.
SrF$_2$ and BaF$_2$ are examined in
Refs.~\cite{Boyer_SSI_1981,Makur_PSSB_1992,Schmalzl_PRB_2007}; as is
the case for $\alpha$-CaF$_2$ \cite{Boyer_PRL_1980}, a critical
softening of a zone-boundary phonon at $X$ is seen at increasing
volumes. Phonon frequencies in stoichiometric Li$_3$OCl have been
discussed in detail in Ref.~\cite{Chen_PRB_2015}, and phonon
instabilities in this compound (such as those shown in Fig.~3(c) of
\cite{Cazorla_Comment_2018}) are attributable to structural phase
transitions which distort the high-symmetry $Pm\overline{3}m$
perovskite structure. This kind of soft phonon mode can actively
promote superionic behaviour, as discussed in
Ref.~\cite{Zhao_JACS_2012}.

AgI has a number of superionic polymorphs \cite{Han_JCP_2014}; without
the benefit of knowing which polymorph CE refer to in their comment,
we restrict our discussion here to $\beta$-AgI and its transition to
the superionic $\alpha$-AgI phase. Ref.~\cite{Buhrer_SSC_1975}
identifies a low-energy optical phonon mode present in $\beta$-AgI
that strongly favours defect creation and thereby drives the
superionic $\beta$-$\alpha$ transition in this
compound. Ref.~\cite{Li_JPCM_2008} provides examples of phonon
modes that critically soften in $\beta$-AgI upon decreasing
volume. That the superionic transition in $\beta$-AgI is first-order
is no barrier to it being driven by either low-energy or soft phonons;
however, the transition will be of the order-disorder, rather than
displacive, type.

Direct experimental evidence for soft-mode behaviour in superionic
materials has been reported in Ref.~\cite{Danilkin_JPSJ_2010} for
superionic Cu$_{2-\delta}$Se. As such, CE's criticisms of our use of
the QHA in \cite{Nelson_PRB_2017} are not backed by experimental
evidence; quite the contrary. Instead, CE's confusion about currently
available experimental results stems from their incorrect assumption
that any soft phonon mode can cause a superionic phase transition.

Lastly, we address the issue of whether the QHA can give
quantitatively accurate superionic transition temperatures. In our
view, this would require (1) thermal expansion to be accurate in the
QHA, and (2) the presence of a harmonic phonon mode (judged to be
linked with the transition) which softens at the same volume at which
the transition occurs. Point (1) was addressed in depth in our
original work \cite{Nelson_PRB_2017}, where we provide a validity
criteria for the QHA based on calculated thermal expansion
coefficients $\alpha$ \cite{Karki-2001,Wentzcovitch-2004}, and which
is depicted in Fig.~\ref{fig:PhaseDiagram}. Point (2) is less obvious,
largely because the exact volume at which a phonon softens is strongly
dependent on exchange-correlation. We expect to discuss it further
in Ref.~\cite{Nelson_PRB_2018}.

\section{\label{Conclusions}Conclusions}
Revisiting our quasiharmonic phase diagram for CaF$_2$ (Fig.~8 of
\cite{Nelson_PRB_2017}), we find it to be completely consistent with
currently available experimental data on high-$PT$ CaF$_2$
(Fig.~\ref{fig:PhaseDiagram}).

The AIMD calculations presented in the comment (Fig.~2 of
\cite{Cazorla_Comment_2018}) are incorrect because CE have not used
appropriately sized simulation cells. CE's description of the
qualitative high-$T$ behaviour of $P\overline{6}2m$-CaF$_2$ is
correspondingly incorrect. When adequately-sized simulation cells are
used, we find that $P\overline{6}2m$-CaF$_2$ undergoes a high-$T$
superionic phase transition (Fig.~\ref{fig:T_2500K_3000K}), as
predicted in our work \cite{Nelson_PRB_2017}. These results highlight
the pitfalls of finite-size effects in molecular dynamics simulations
and the importance of converging such simulations with respect to
system size.

The QHA correctly predicts superionic phase transitions in
$\alpha$-CaF$_2$ and in $P\overline{6}2m$-CaF$_2$. The discussion in
Sec.~\ref{Predict}, and references therein, makes it clear that this
is a very general conclusion, and demonstrates the utility of the QHA
in identifying superionic behaviour.

\end{document}